\documentclass[aps,prb,a4paper,10pt,twocolumn,bibnotes,floatfix,superscriptaddress,preprintnumbers,longbibliography]{revtex4-1}
\usepackage{float}
\usepackage{dcolumn,graphicx,color,booktabs,microtype,afterpage}
\usepackage{amsmath}
\graphicspath{{./}{figure/}}
\usepackage[charter,greekuppercase=italicized]{mathdesign}
\usepackage{sidecap}
\usepackage[mathlines]{lineno}
\renewcommand{\tablename}{Table}
\makeatletter\renewcommand{\fnum@figure}[1]{\figurename~\thefigure.~}\makeatother
\makeatletter\renewcommand{\fnum@table}[1]{\tablename~\thetable.}\makeatother

\newcount\hh \newcount\mm
\hh=\time \divide\hh by 60
\mm=\hh \multiply\mm by 60 \mm=-\mm
\advance\mm by \time
\def\now{\number\hh:\ifnum\mm<10{}0\fi\number\mm}

\usepackage[colorlinks,plainpages=false,linkcolor=blue,urlcolor=blue,citecolor=blue,pdfpagemode=UseNone,pdfstartview=FitBH]{hyperref}

\newcommand{\tcr}[1]{\textcolor{black}{#1}}

\newcommand{\NIO}{Nb$_5$Ir$_3$O}
\newcommand{\NIPO}{Nb$_5$Ir$_{1.4}$Pt$_{1.6}$O}

\begin{document}

\makeatletter\renewcommand{\ps@plain}{%
\def\@evenhead{\hfill\itshape\rightmark}%
\def\@oddhead{\itshape\leftmark\hfill}%
\renewcommand{\@evenfoot}{\hfill\small{--~\thepage~--}\hfill}%
\renewcommand{\@oddfoot}{\hfill\small{--~\thepage~--}\hfill}%
}\makeatother\pagestyle{plain}

\preprint{\textit{Preprint: \today, \now.}} 

%
\title{Multiple- to single-gap superconductivity crossover in Nb$_5$Ir$_{3-x}$Pt$_x$O alloys}
\author{Y.\ Xu}\email[Corresponding author: ]{yangxu@physik.uzh.ch}
\affiliation{Physik-Institut, Universität Z\"{u}rich, Winterthurerstrasse 190, CH-8057 Z\"{u}rich, Switzerland}
%
%
\author{S.\ J\"{o}hr}
\affiliation{Physik-Institut, Universität Z\"{u}rich, Winterthurerstrasse 190, CH-8057 Z\"{u}rich, Switzerland}
%
\author{L.\ Das}
\affiliation{Physik-Institut, Universität Z\"{u}rich, Winterthurerstrasse 190, CH-8057 Z\"{u}rich, Switzerland}

%
\author{J.\ Kitagawa}
\affiliation{Department of Electrical Engineering, Faculty of Engineering, Fukuoka Institute of Technology, 3-30-1 Wajiro-higashi, Higashi-ku, Fukuoka, 811-0295, Japan}
%
%
\author{M.\ Medarde}
\affiliation{Laboratory for Multiscale Materials Experiments, Paul Scherrer Institut, Villigen CH-5232, Switzerland}
%
%
\author{T.\ Shiroka}
\affiliation{Laboratorium f\"ur Festk\"orperphysik, ETH Z\"urich, CH-8093 Zurich, Switzerland}
\affiliation{Paul Scherrer Institut, CH-5232 Villigen PSI, Switzerland}
%
\author{J.\ Chang}
\affiliation{Physik-Institut, Universität Z\"{u}rich, Winterthurerstrasse 190, CH-8057 Z\"{u}rich, Switzerland}

\author{T.\ Shang}\email[Corresponding author: ]{tian.shang@psi.ch}
\affiliation{Physik-Institut, Universität Z\"{u}rich, Winterthurerstrasse 190, CH-8057 Z\"{u}rich, Switzerland}
\affiliation{Laboratory for Multiscale Materials Experiments, Paul Scherrer Institut, Villigen CH-5232, Switzerland}

\begin{abstract}
By using mostly  
the muon-spin rotation/relaxation ($\mu$SR) technique,
we investigate the superconductivity (SC) of Nb$_5$Ir$_{3-x}$Pt$_x$O ($x = 0$ and 1.6) alloys, with 
$T_c = 10.5$\,K and 9.1\,K, respectively. 
At a macroscopic level, their superconductivity 
was studied by electrical resistivity, magnetization, and specific-heat measurements. 
In both compounds, the electronic specific heat and the low-temperature superfluid density data
suggest a nodeless SC.
The superconducting gap value 
and the specific heat discontinuity at $T_c$ 
are larger than that expected from the Bardeen-Cooper-Schrieffer theory in 
the weak-coupling regime, indicating strong-coupling superconductivity in the Nb$_5$Ir$_{3-x}$Pt$_x$O family.    
In Nb$_5$Ir$_3$O, multigap SC is evidenced by the field dependence of the electronic  
specific heat coefficient and the superconducting Gaussian relaxation rate, as well as by the temperature 
dependence of the upper critical field. Pt substitution suppresses one of the gaps, and Nb$_5$Ir$_{1.4}$Pt$_{1.6}$O 
becomes a single-gap superconductor. 
By combining our extensive experimental results, 
we provide evidence for a multiple- to single-gap SC crossover in the Nb$_5$Ir$_{3-x}$Pt$_x$O family.  

\end{abstract}

\maketitle\enlargethispage{3pt}

\vspace{-5pt}
\section{\label{sec:Introduction}Introduction}\enlargethispage{8pt}

\begin{figure}[th]
	\centering
	\includegraphics[width=0.49\textwidth,angle=0]{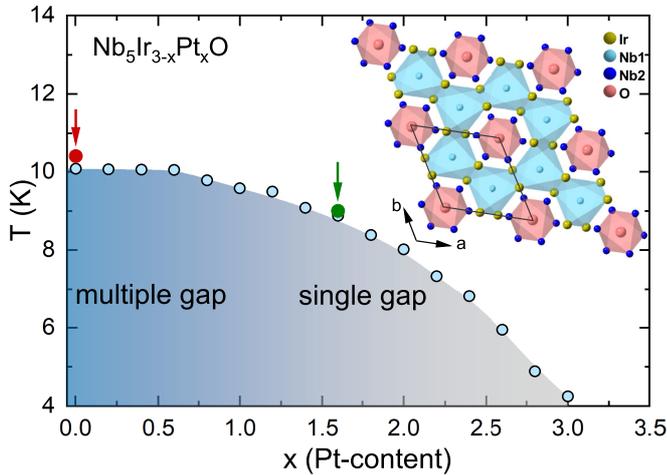}
	\vspace{-2ex}%
	\caption{\label{fig:phase_diag}The $T$-$x$ phase diagram of Nb$_5$Ir$_{3-x}$Pt$_x$O. The arrows indicate the two samples studied here. The inset shows the crystal structure of Nb$_5$Ir$_3$O viewed along the $c$-axis (solid lines mark the unit cell). A crossover from multiple- to single-gap SC with increasing Pt-content has been proposed. Data were adopted from Ref.~\onlinecite{Kitagawa2019}.}
\end{figure}

The $A_5B_3$ compound family, where $A$ is a transition or rare-earth metal, and $B$ a (post)-transition metal or a metalloid element, consists of more than five hundreds 
compounds with three  
distinct crystal structures. They are orthorhombic Yb$_5$Sb$_3$- ($Pnma$, No.\ 62), tetragonal Cr$_5$B$_3$- ($I4/mcm$, No.\ 140), and hexagonal Mn$_5$Si$_3$-type ($P6_3/mcm$, No.\ 193) structures. The latter possesses an interstitial 2$b$ site, 
allowing the intercalation of light atoms, e.g., oxygen, boron, and carbon, to engineer the band topology. 
The ordered variant of Mn$_5$Si$_3$-type structure is also known as Ti$_5$Ga$_4$- or Hf$_5$CuSn$_3$-type structure. 
To the latter belongs also   
Nb$_5$Ir$_3$O, whose crystal stucture is presented in the inset of Fig.~\ref{fig:phase_diag}. 
Superconductivity (SC) has been reported to occur in several
Ti$_5$Ga$_4$-type compounds, including Nb$_5$Ir$_3$O~\cite{Zhang2017}, (Nb,Zr)$_5$Pt$_3$O~\cite{Cort1982,Hamamoto2018}, Nb$_5$Ge$_3$C$_{0.3}$~\cite{Bortolozo2012}, or Zr$_5$Pt$_3$C$_x$~\cite{Renosto2018}, with the highest superconducting transition temperature $T_c$ reaching $\sim$ 15\,K.
Interestingly, specific heat 
and penetration depth 
results suggest a nodal superconducting gap in Zr$_5$Pt$_3$C$_x$ 
and possibly unconventional SC~\cite{Renosto2018}. 

The parent compound Nb$_5$Ir$_3$ consists of mixed tetragonal and hexagonal 
phases, both showing superconducting behavior below $T_c$ = 2.8 and 9.4\,K, respectively. The gradual intercalation of oxygen suppresses 
the tetragonal phase, 
making the hexagonal phase the dominant one. Accordingly, the $T_c$ value 
increases linearly with the interstitial oxygen content, to reach 10.5\,K 
in the purely hexagonal Nb$_5$Ir$_3$O~\cite{Zhang2017}. Nb$_5$Ir$_3$ exhibits an unusual interplay of electride and SC states, whereas only SC remains in
Nb$_5$Ir$_3$O~\cite{Zhang2017}.
Density functional theory (DFT) calculations indicate Nb$_5$Ir$_3$
to be a multiband metal, whose density of states (DOS) at
the Fermi level is dominated
by the Nb 4$d$- and Ir 5$d$-orbitals~\cite{Zhang2017}.
The increase of $T_c$ in Nb$_5$Ir$_3$O$_x$ is most likely attributed to 
the enhanced electron-phonon coupling strength or to an increased 
DOS at the Fermi level (with extra Nb-4$d$ and O-2$p$ contributions)~\cite{Zhang2017}.
On the other hand, by applying external pressure, $T_c$ decreases monotonically from 10.5\,K at ambient pressure to 9.5\,K at 13\,GPa~\cite{Zhang2017,Wang2019}.
Unlike the Nb$_5$Ir$_3$O$_x$ case, in  Zr$_5$Pt$_3$ the addition of 
oxygen reduces the $T_c$ value from 6.4\,K to 3.2\,K (in Zr$_5$Pt$_3$O$_{0.6}$)~\cite{Hamamoto2018}.
Similarly, in Zr$_5$Sb$_3$, the addition of oxygen reduces the 
$T_c$, with Zr$_5$Sb$_3$O 
being a normal metal down to 1.8\,K~\cite{Lv2013}.

Although the superconductivity of several Mn$_5$Si$_3$- or Ti$_5$Ga$_4$-types of compounds 
has been studied by magnetic and transport measurements and, in many cases,
electronic band-structure calculations are available,   
the microscopic nature of their superconducting phase remains largely unexplored.  
In Nb$_5$Ir$_3$O, the low-$T$ electronic specific heat data suggest a nodeless SC with multiple gaps~\cite{Zhang2017,Wang2019}. While, in the Nb$_5$Pt$_3$O case, 
the electronic specific heat shows a sin\-gle\--ex\-po\-nen\-tial temperature dependence below $T_c$,   
more consistent with a single-gap superconductivity~\cite{Cort1982}.
In Nb$_5$Ir$_{3-x}$Pt$_x$O, Pt substitution increases the 
$a$-axis lattice constant, while reducing the $c$-axis. 
As shown in Fig.~\ref{fig:phase_diag}, $T_c$ is almost independent of Pt-content for $x \le 0.5$, but it starts to decrease continuously for $x > 0.5$, reaching 4.3\,K in Nb$_5$Pt$_3$O~\cite{Hamamoto2018}. 
As hinted by specific heat data, one expects 
a crossover from multiple- to single gap SC in Nb$_5$Ir$_{3-x}$Pt$_x$O.

To provide further evidence for such a crossover, 
we initiated an extensive study of the superconducting properties of Nb$_5$Ir$_{3-x}$Pt$_x$O for $x$ = 0 and 1.6, two representative cases in the multiple- and single gap regions (see arrows in Fig.~\ref{fig:phase_diag}), 
probing them at both the macroscopic 
and microscopic level. Both compounds were found to be fully-gapped superconductors, with Nb$_5$Ir$_3$O showing multiple gaps and Nb$_5$Ir$_{1.4}$Pt$_{1.6}$O being a single gap superconductor.

%

%
\section{Experimental details\label{sec:details}}\enlargethispage{8pt}

Polycrystalline samples of Nb$_5$Ir$_{3-x}$Pt$_x$O were prepared by the 
arc-melting method (the full details are reported 
in Ref.~\onlinecite{Kitagawa2019}). 
The magnetic susceptibility, electrical resistivity, and specific heat measurements 
were performed on a 7-T Quantum Design Magnetic Property Measurement 
System (MPMS-7) and a 9-T Physical Property Measurement System (PPMS-9). 
The muon-spin rotation/relaxation ($\mu$SR) experiments were carried out at the general purpose surface-muon (GPS) spectrometer at the Swiss muon source (S$\mu$S) at Paul Scherrer Institut, 
Villigen, Switzerland~\cite{Amato2017}. Both transverse-field (TF-) and zero-field (ZF-) $\mu$SR measurements were performed.
The $\mu$SR data were analyzed using the \texttt{musrfit} software package~\cite{Suter2012}.

\section{\label{sec:results} Results and discussion}\enlargethispage{8pt}
\subsection{\label{ssec:structure} Magnetization}

\begin{figure}[th]
	\centering
	\includegraphics[width=0.44\textwidth,angle=0]{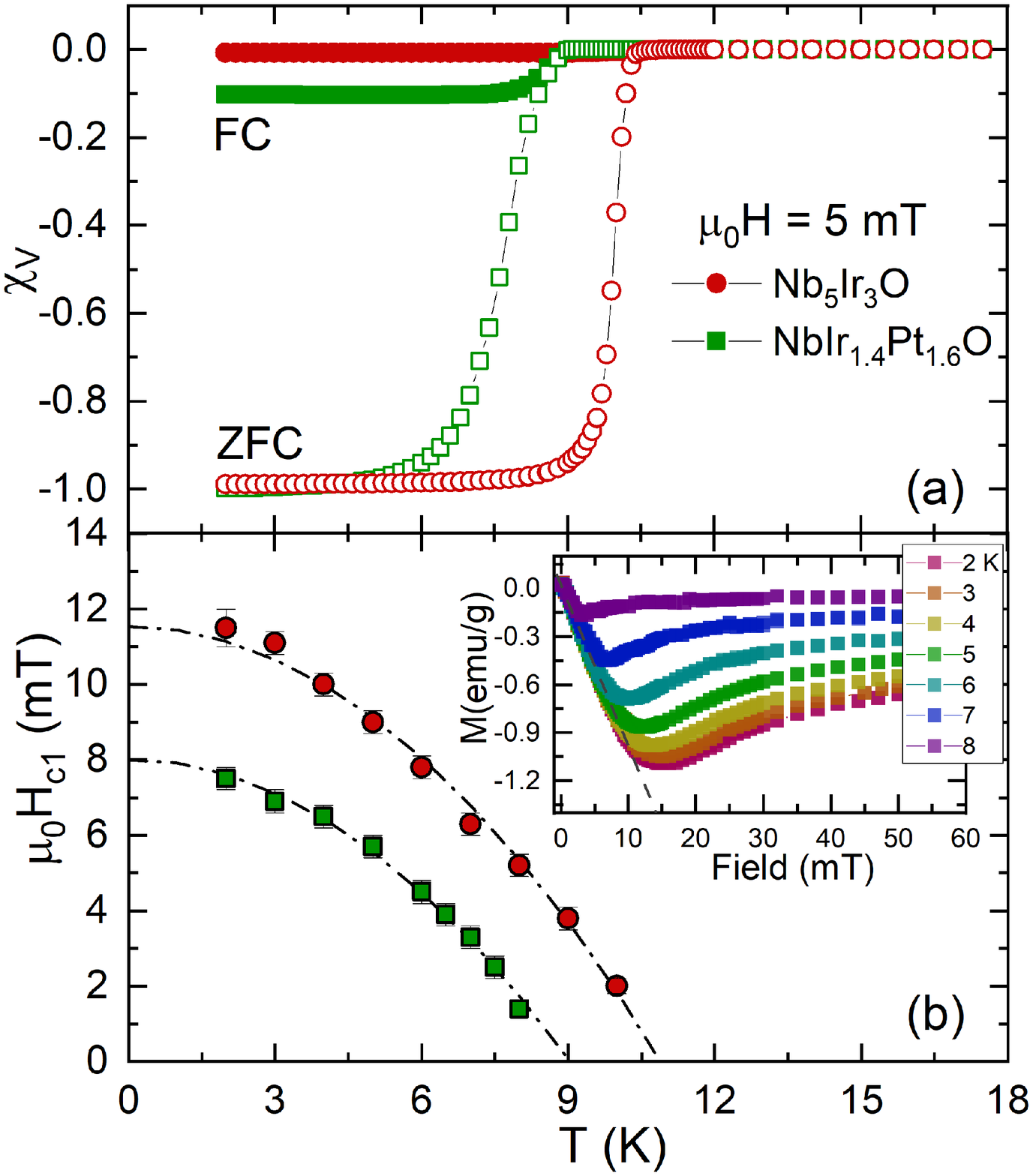}
	\caption{\label{fig:Chi}(a) Temperature dependence of the magnetic susceptibility 
		of Nb$_5$Ir$_3$O  and Nb$_5$Ir$_{1.4}$Pt$_{1.6}$O.
		(b) Estimated lower critical field $\mu_{0}H_{c1}$ vs.
		temperature. The dash-dotted lines are fits to 
		$\mu_{0}H_{c1}(T) =\mu_{0}H_{c1}(0)[1-(T/T_{c})^2]$. 
		The inset shows the field-dependent magnetization $M(H)$ recorded 
		at various temperatures up to $T_c$ for \NIPO. For each temperature, the 
		lower critical field $\mu_{0}H_{c1}$ was determined as the value 
		where $M(H)$ starts deviating from linearity (dashed line). The magnetic susceptibilities were corrected by using the demagnetization
		factor obtained from the field-dependent magnetization at 2\,K (base temperature).}
\end{figure}
%

The SC of  Nb$_5$Ir$_{3-x}$Pt$_x$O ($x$ = 0, 1.6) was first characterized by magnetic susceptibility measurements, carried out in a 5-mT field, using both field-cooled (FC) and zero-field-cooled (ZFC) protocols. 
As shown in Fig.~\ref{fig:Chi}(a), the ZFC-susceptibility, corrected to account for the demagnetization factor, indicates bulk SC below $T_c$ = 10.5\,K and 9.1\,K for Nb$_5$Ir$_3$O and Nb$_5$Ir$_{1.4}$Pt$_{1.6}$O, respectively.
The well separated ZFC- and FC-susceptibilities indicate strong flux-line pinning 
across the Nb$_5$Ir$_{3-x}$Pt$_x$O series.  
To perform TF-$\mu$SR measurements on superconductors, the applied magnetic field should  
exceed the lower critical field $\mu_{0}H_{c1}$, so that the additional field-distribution broadening due to the flux-line lattice (FLL) can be quantified from the muon-spin relaxation rate. 
To determine $\mu_{0}H_{c1}$, the field-dependent magnetization $M(H)$ was measured at various temperatures up to $T_c$. As an example, the $M(H)$ data for Nb$_5$Ir$_{1.4}$Pt$_{1.6}$O are plotted in the inset of  Fig.~\ref{fig:Chi}(b), with Nb$_5$Ir$_3$O showing similar features. The estimated  $\mu_{0} H_{c1}$ values are summarized in Fig.~\ref{fig:Chi}(b) as a function of temperature for both samples. 
The dash-dotted lines represent fits to $\mu_{0}H_{c1}(T) = \mu_{0}H_{c1}(0)[1-(T/T_{c})^2]$ and yield lower critical fields 
of 11.5(2) and 8.0(1)\,mT for Nb$_5$Ir$_3$O and Nb$_5$Ir$_{1.4}$Pt$_{1.6}$O, respectively.  

\subsection{\label{ssec:critical_field}  Upper critical field}  
%

\begin{figure}[th]
	\centering
	\includegraphics[width=0.47\textwidth,angle=0]{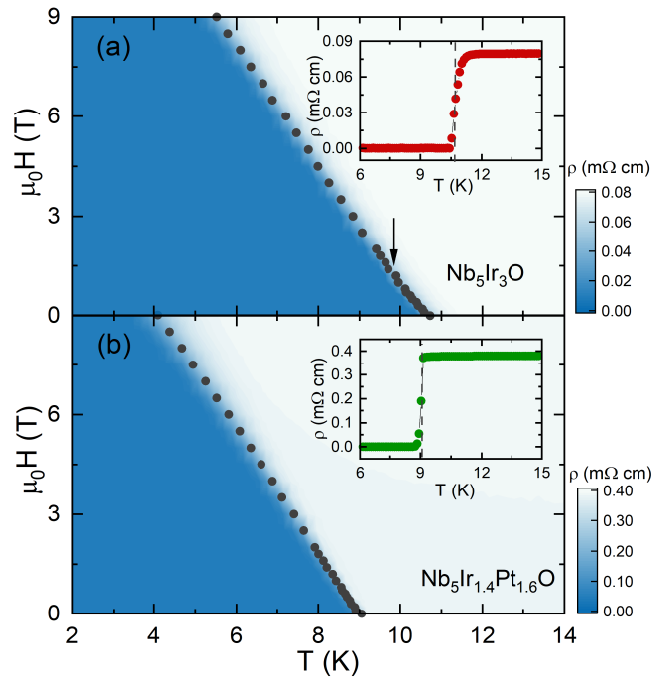}
	\vspace{-2ex}%
	\caption{\label{fig:rho}
		\tcr{Contour plots showing the electrical resistivity vs.\ temperature (down to 2\,K) and magnetic field (up to 9\,T) for Nb$_5$Ir$_3$O (a) 
		and Nb$_5$Ir$_{1.4}$Pt$_{1.6}$O (b)}. Color hues represent the absolute value of electrical resistivity.
		The symbols indicate the critical temperatures $T_c$, as determined from the middle of superconducting transition (see dashed lines in the insets). Insets: In both cases, the zero-field electrical resistivity shows a sharp transition.
		The arrow in (a) indicates a change of slope (see also 
		inset in Fig.~\ref{fig:Hc2_twoband}).}
\end{figure}
\begin{figure}[!th]
	\centering
	\includegraphics[width=0.48\textwidth,angle=0]{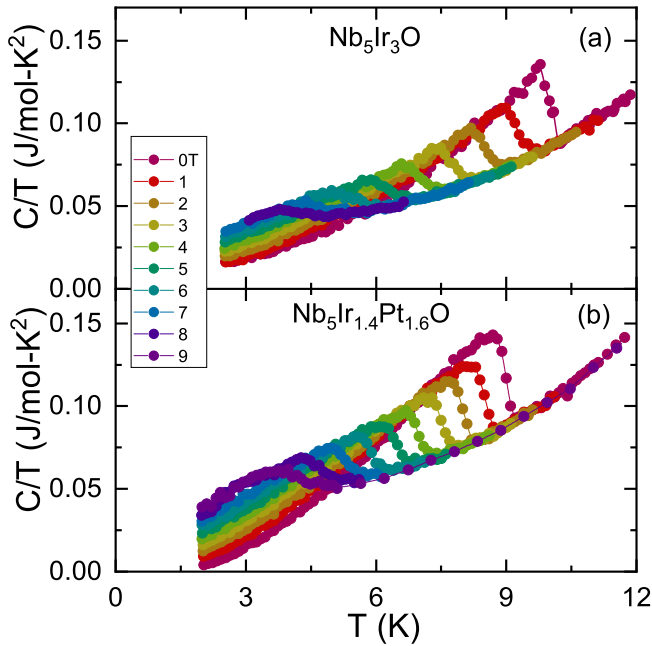}
	\vspace{-2ex}%
	\caption{\label{fig:Cp1}Temperature dependence of the specific heat of Nb$_5$Ir$_3$O (a) and Nb$_5$Ir$_{1.4}$Pt$_{1.6}$O (b) measured under various magnetic fields up to 9\,T. The data were collected upon sample heating.} 
\end{figure}

%
The upper critical field $\mu_0$$H_{c2}$ of Nb$_5$Ir$_{3-x}$Pt$_x$O was determined 
from the measurements of electrical resistivity $\rho(T, H)$, specific heat $C(T, H)/T$, and magnetization $M(T, H)$ 
under various applied magnetic fields up to 9\,T. Here we show $\rho(T, H)$ and $C(T, H)/T$ for both samples in Fig.~\ref{fig:rho} and Fig.~\ref{fig:Cp1}, respectively. 
Upon applying a magnetic field, the superconducting transition,
as detected using either $\rho(T)$ 
or $C(T)$/$T$, shifts towards lower temperatures. 
The insets of Fig.~\ref{fig:rho} show the enlarged plot of zero-field electrical resistivity. Nb$_5$Ir$_3$O exhibits an onset of the SC at $T_c^\mathrm{onset}$ = 10.9\,K, and its resistivity drops to zero at $T_c^\mathrm{zero} = 10.5$\,K. For Nb$_5$Ir$_{1.4}$Pt$_{1.6}$O, $T_c^\mathrm{onset}$ = 9.1\,K and $T_c^\mathrm{zero} = 8.9$\,K. 
The sharp superconducting transitions ($\Delta T$ $\sim$ 0.2-0.4\,K) in 
zero field, confirmed also by specific heat data (see Fig.~\ref{fig:Cp1}), indicate the good quality of the samples.  

\begin{figure}[!htp]
	\centering
	\includegraphics[width=0.5\textwidth,angle= 0]{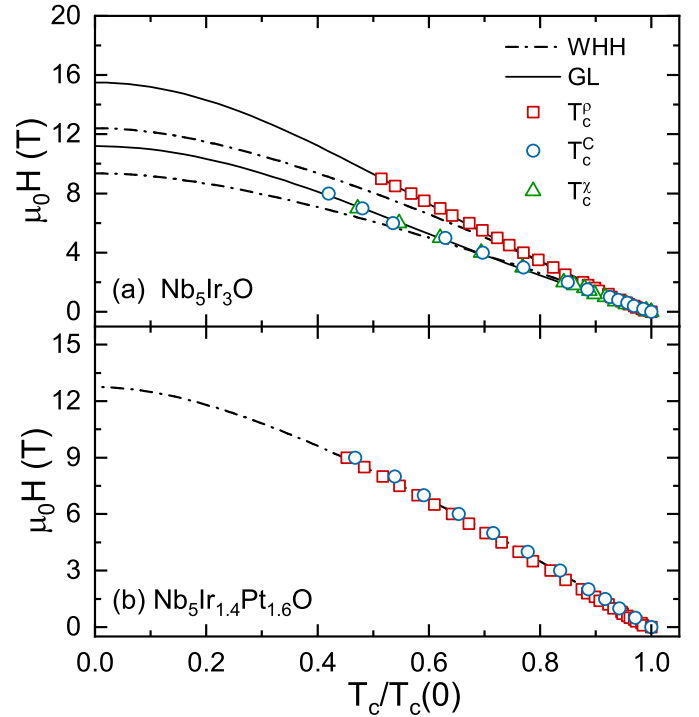}
	\caption{\label{fig:Hc2_determ}Upper critical field $\mu_{0}H_{c2}$ vs.\ 
		reduced transition temperature $T_c/T_c(0)$ for Nb$_5$Ir$_3$O (a) and Nb$_5$Ir$_{1.4}$Pt$_{1.6}$O (b). The $T_c$ values 
		were determined from temperature-dependent electrical resistivity $\rho(T, H)$, magnetization $M(T, H)$, 
		and specific heat $C(T,H)/T$. For
		$\rho(T,H)$ measurements, $T_c$ is defined as the middle of the superconducting transition. 
		Two different models, including GL- (solid lines) and WHH models (dash-dotted lines), were used 
		to analyze the $\mu_{0}H_{c2}(T)$ data. For the WHH model, the spin-orbit scattering was neglected.}
\end{figure}

The determined $\mu_0H_{c2}$ values as a function of the reduced temperature 
$T_c$/$T_c$(0) are summarized in Fig.~\ref{fig:Hc2_determ}. Here $T_c$(0) is the transition temperature in
zero field. 
For Nb$_5$Ir$_{1.4}$Pt$_{1.6}$O, the $\mu_0 H_{c2}$ values determined using different techniques are highly consistent. Conversely, for Nb$_5$Ir$_3$O, 
the datasets agree well only at low fields (below 2\,T),  
since at higher fields the transition temperatures determined from $\rho(T)$ data are 
 systematically higher than those 
derived from $C(T)/T$. The surface/filamentary SC above bulk $T_c$ might cause the different $T_c$ values, but why it shows up only in Nb$_5$Ir$_3$O is not clear yet. 
The temperature dependence of $\mu_0 H_{c2}(T)$ was analyzed by means 
of Ginzburg-Landau- (GL)~\cite{Zhu2008} and Werthamer-Helfand-Hohenberg (WHH) models~\cite{Werthamer1966}. 
For Nb$_5$Ir$_3$O, the better agreement of the GL model with the data is clearly seen in Fig.~\ref{fig:Hc2_determ}(a).
At low fields, both GL and WHH models reproduce very well the experimental data. However, at higher fields, 
the WHH model deviates significantly from the experimental data, giving underestimated values of $\mu_0 H_{c2}^\mathrm{WHH}(0) =$ 9.3(1)\,T and 12.4(1)\,T for $C(T,H)/T$
and $\rho(T,H)$, respectively.
In contrast, the GL model fits the data over the entire field range, providing $\mu_0 H_{c2}^\mathrm{GL}(0) =$ 11.2(1)\,T ($C/T$) and 15.5(1)\,T ($\rho$). In the case of Nb$_5$Ir$_{1.4}$Pt$_{1.6}$O, $\mu_0 H_{c2}(T)$ are reproduced very well by the WHH model, which yields $\mu_0 H_{c2}^\mathrm{WHH}(0) = $ 12.7(1)\,T.

%

The superconducting coherence length $\xi$ can be calculated from $\xi$ =  $\sqrt{\Phi_0/2\pi\,H_{c2}}$, 
where $\Phi_0 = 2.07 \times 10^{-3}$\,T~$\mu$m$^{2}$ is the magnetic flux quantum. 
With a bulk $\mu_{0}H_{c2}(0) = 11.2(1)$ and 12.7(1)\,T, the calculated $\xi(0)$ is 5.4(1) and 5.1(1)\,nm for Nb$_5$Ir$_3$O and Nb$_5$Ir$_{1.4}$Pt$_{1.6}$O, respectively.  
The lower critical field $\mu_{0}H_{c1}$ is related to the magnetic penetration 
depth $\lambda$ and the coherence length $\xi$ via $\mu_{0}H_{c1} = (\Phi_0 /4 \pi \lambda^2)[$ln$(\kappa)+ 0.5]$, 
where $\kappa$ = $\lambda$/$\xi$ is the GL parameter~\cite{Brandt2003}.
By using $\mu_{0}H_{c1} = 11.5(2)$\,mT [8.0(1)\,mT] and $\mu_{0}H_{c2} = 11.2(1)$\,T [12.7(1)\,T], 
the resulting magnetic penetration depth $\lambda_\mathrm{GL}$ = 249(3)\,nm [308(3)\,nm] for Nb$_5$Ir$_3$O (Nb$_5$Ir$_{1.4}$Pt$_{1.6}$O), 
is comparable to 230(2)\,nm [314(2)\,nm], the experimental values evaluated from TF-$\mu$SR data (see Sec.~\ref{ssec:TF_muSR}). 
A large GL parameter $\kappa \sim 50-60$, clearly indicates that Nb$_5$Ir$_{3-x}$Pt$_x$O are type-II superconductors.

\subsection{\label{ssec:Cp_zero} Zero-field specific heat}
%
%

\begin{figure}[!thp]
	\centering
	\includegraphics[width=0.48\textwidth,angle=0]{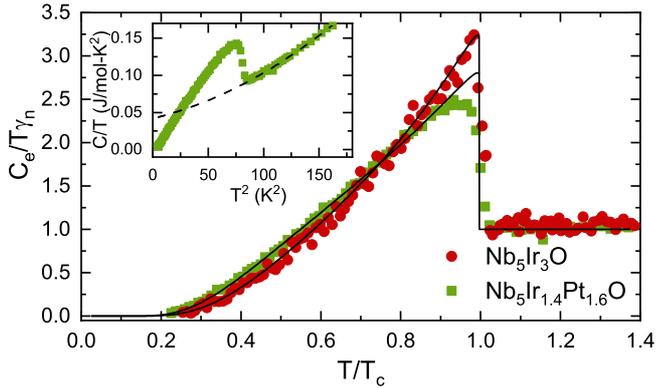}
	\vspace{-2ex}%
	\caption{\label{fig:Cp2} Normalized electronic specific heat 
		$C_\mathrm{e}/\gamma_n T$ of Nb$_5$Ir$_3$O and Nb$_5$Ir$_{1.4}$Pt$_{1.6}$O
		as a function of the reduced temperature $T$/$T_c$. 
		Inset: specific heat $C/T$ vs.\ $T^2$ for Nb$_5$Ir$_{1.4}$Pt$_{1.6}$O; 
		the dashed-line is a fit  
		to $C/T = \gamma_n + \beta T^2 + \delta T^4$ for $T > T_{c}$. 
		The solid lines in the main panel represent the electronic 
		specific heat calculated by considering a fully-gapped $s$-wave model.}
\end{figure}

%
The zero-field specific heat data in the low-$T$ region (Fig.~\ref{fig:Cp2}) show a sharp jump at $T_c$ 
again indicating a bulk superconducting transition and a good sample quality. 
As shown in the inset, the normal-state specific heat data of 
Nb$_5$Ir$_{3-x}$Pt$_x$O were fitted to $C/T = \gamma_\mathrm{n} + \beta T^2 + \delta T^4$, 
where $\gamma_\mathrm{n}$ is the normal-state electronic specific heat coefficient, 
while the two other terms account for the phonon contribution to the 
specific heat. The derived values are $\gamma_\mathrm{n} = 37(5)$\,mJ/mol-K$^2$, 
$\beta = 0.32(9)$\,mJ/mol-K$^4$ and $\delta = 1.7(3)$\,$\mu$J/mol-K$^6$ for Nb$_5$Ir$_3$O, 
and  $\gamma_\mathrm{n} = 42(6)$\,mJ/mol-K$^2$, 
$\beta = 0.36(13)$\,mJ/mol-K$^4$ and $\delta = 2.6(5)$\,$\mu$J/mol-K$^6$ for Nb$_5$Ir$_{1.4}$Pt$_{1.6}$O, respectively. Such large $\gamma$ values suggest a relatively large effective electron mass (see Table~\ref{tab:parameter}) and also strong electronic correlations in Nb$_5$Ir$_{3-x}$Pt$_x$O.
The Debye temperature can be estimated using  $\Theta_\mathrm{D} = (12\pi^4\,Rn/5\beta)^{1/3}$, where $R$ = 8.314\,J/mol-K is the molar gas constant and $n$ = 9
is the number of atoms per formula unit, giving $\Theta_\mathrm{D}$ = 380(7)  and 365(8)\,K for Nb$_5$Ir$_3$O and  Nb$_5$Ir$_{1.4}$Pt$_{1.6}$O. 
The density of states at the Fermi level $N(\epsilon_\mathrm{F})$ 
is evaluated to be
$N(\epsilon_\mathrm{F}) = 3\gamma_\mathrm{n}/(\pi^2 k_\mathrm{B}^2) = 16(2)$ (Nb$_5$Ir$_3$O) and 18(2)\,states/eV-f.u. (Nb$_5$Ir$_{1.4}$Pt$_{1.6}$O)~\cite{Kittel2005}, where 
$k_\mathrm{B}$ is the Boltzmann constant. Both values are comparable to those from the electronic band structure calculations~\cite{Zhang2017}. 
The electron-phonon coupling 
constant $\lambda_\mathrm{ep}$, a measure of the attractive interaction 
between electrons due to phonons, was estimated using the semi-empirical 
McMillan formula~\cite{McMillan1968}:
\begin{equation}
\label{eq:lambda_ep}
\lambda_\mathrm{ep}=\frac{1.04+\mu^{\star}\,\mathrm{ln}(\Theta_\mathrm{D}/1.45\,T_c)}{(1-0.62\,\mu^{\star})\mathrm{ln}(\Theta_\mathrm{D}/1.45\,T_c)-1.04}.
\end{equation}
The Coulomb pseudo-potential $\mu^{\star}$ is material-specific, 
typically lying in the 0.1 $\leq \mu^{\star} \leq$ 0.15 range. 
Here, $\mu^{\star}$ was fixed to 0.13, a typical 
value for metallic samples~\cite{McMillan1968}. From Eq.~\eqref{eq:lambda_ep} we obtain 
$\lambda_\mathrm{ep} = 0.8(2)$ (Nb$_5$Ir$_3$O) and 0.7(3)  
(Nb$_5$Ir$_{1.4}$Pt$_{1.6}$O), 
which classifies both of them as relatively strongly coupled superconductors.  
This is consistent with previous results~\cite{Zhang2017,Wang2019}, \tcr{and compatible with other strongly-coupled superconductors, as e.g, Ba$_{1-x}$K$_x$BiO$_3$ ($\lambda_\mathrm{ep} \sim 1$), or W$_3$Al$_2$C ($\lambda_\mathrm{ep} \sim 0.78$)~\cite{Wen2018,Shang2020,Ying2019}.} 

The electronic specific heat is obtained by subtracting the phonon contribution from the total specific heat. The main panel in Fig.~\ref{fig:Cp2} shows the temperature dependence of $C_\mathrm{e} / \gamma_\mathrm{n} T$, from which one can evaluate 
the discontinuity at $T_c$ to be $\Delta$$C$/$\gamma$$T_c$ = 2.24(7) for Nb$_5$Ir$_3$O and 1.50(5) for Nb$_5$Ir$_{1.4}$Pt$_{1.6}$O, both larger than the weak-coupling Bardeen-Cooper-Schrieffer (BCS) value of 1.43. The temperature evolution of the SC-related contribution to the entropy 
can be 
calculated from the BCS expression~\cite{Tinkham1996}:
\begin{equation}
\label{eq:entropy}
S(T) = -\frac{6\gamma_\mathrm{n}}{\pi^2 k_\mathrm{B}} \int^{\infty}_0 [f\mathrm{ln}f+(1-f)\mathrm{ln}(1-f)]\,\mathrm{d}\epsilon,
\end{equation}
where $f = (1+e^{E/k_\mathrm{B}T})^{-1}$ is the Fermi distribution and 
$E(\epsilon) = \sqrt{\epsilon^2 + \Delta^2(T)}$ is the excitation energy 
of quasiparticles, with $\epsilon$ the electron energy measured relative 
to the chemical potential (Fermi energy)~\cite{Padamsee1973,Tinkham1996}. 
Here $\Delta(T) = \Delta_0 \mathrm{tanh} \{ 1.82[1.018(T_\mathrm{c}/T-1)]^{0.51} \}$ 
\cite{Carrington2003}, with $\Delta_0$ the zero temperature superconducting gap value. 
The electronic specific heat in the superconducting
state can be calculated from $C_\mathrm{e} =T \frac{dS}{dT}$. 
The solid lines in Fig.~\ref{fig:Cp2} are fits to the $s$-wave model with a single gap $\Delta_0 = 1.89(2)$ and 1.53(1)\,meV for Nb$_5$Ir$_3$O and
Nb$_5$Ir$_{1.4}$Pt$_{1.6}$O, respectively. 
Both gap values are consistent with the TF-$\mu$SR results (see Fig.~\ref{fig:lambda} below) and are 
larger than the standard BCS value ($\Delta_0 \sim 1.73$\,k$_\mathrm{B}T_c \sim 1.4$\,meV) in the weak-coupling limit. This indicates strongly-coupled 
superconducting pairs in Nb$_5$Ir$_{3-x}$Pt$_x$O, consistent with the 
large $\lambda_\mathrm{ep}$ values. 
In previous studies, two superconducting gaps were required to describe the zero-field electronic specific heat at $T$ < 1/3$T_c$~\cite{Zhang2017,Wang2019}. 
Note, however, that the extra gap value was significantly smaller 
($< 0.02$\,meV) than the large gap and that it accounts for less than 10\% 
of the total weights~\cite{Zhang2017,Wang2019}. As such, it can easily 
be influenced by
disorder or oxygen content.  
In the case of our samples, a single gap could reproduce the electronic specific heat data very well over 
the whole temperature range.
However, the field-dependent electronic specific heat coefficient, the TF-$\mu$SR relaxation rates, and the temperature-dependent upper critical fields, all suggest multigap features in Nb$_5$Ir$_{3}$O, while Nb$_5$Ir$_{1.4}$Pt$_{1.6}$O is a single gap superconductor (see details in Sec.~\ref{ssec:twoband}).   

\begin{figure}[!thp]
	\centering
	\includegraphics[width=0.48\textwidth,angle= 0]{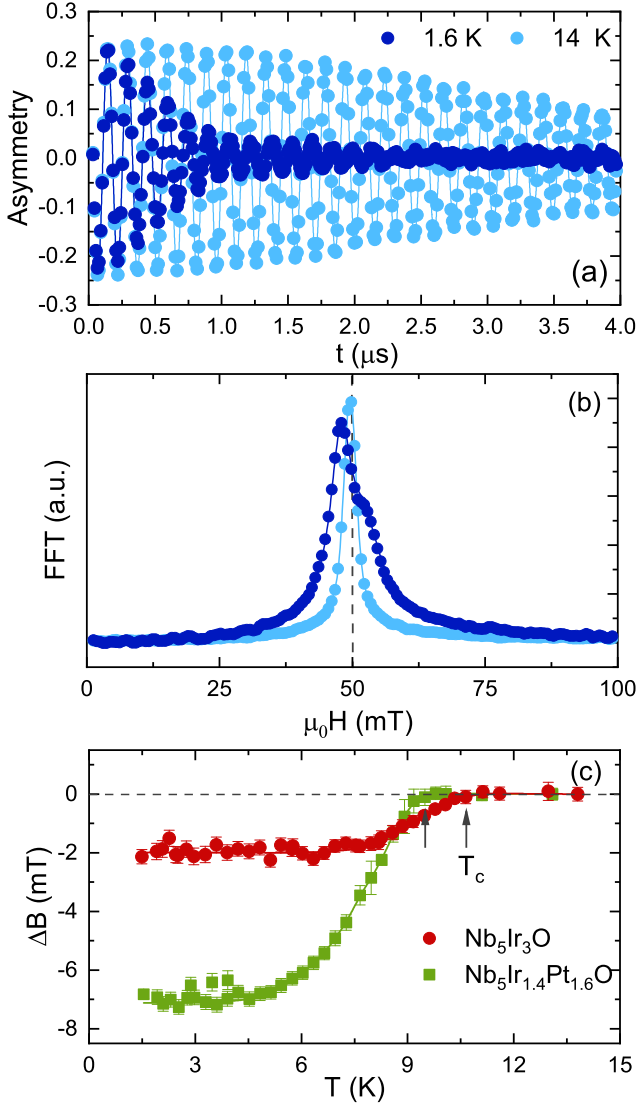}
	\caption{\label{fig:TF-muSR_T}(a) TF-$\mu$SR time spectra collected 
		at temperatures below (1.6\,K) and above $T_c$ (14 \,K) in an applied field of 50\,mT for Nb$_5$Ir$_3$O.   
		The fast Fourier transform of the TF-$\mu$SR time spectra  are shown in (b). 
		Note that Nb$_5$Ir$_{1.4}$Pt$_{1.6}$O exhibits similar spectra.
		Solid lines in (a) and (b) are fits to Eq.~\eqref{eq:TF_muSR} using two Gaussian relaxations.  
		The dashed vertical line indicates the applied magnetic field, showing a clear diamagnetic shift.
		(c) Diamagnetic shift ($\Delta B = \langle B \rangle - B_\mathrm{appl.}$) vs.\ temperature for Nb$_5$Ir$_{3}$O and Nb$_5$Ir$_{1.4}$Pt$_{1.6}$O. }
\end{figure}

\subsection{\label{ssec:TF_muSR} Transverse-field \texorpdfstring{$\mu$SR}{MuSR}}

\begin{figure}[!thp]
	\centering
	\includegraphics[width=0.49\textwidth,angle= 0]{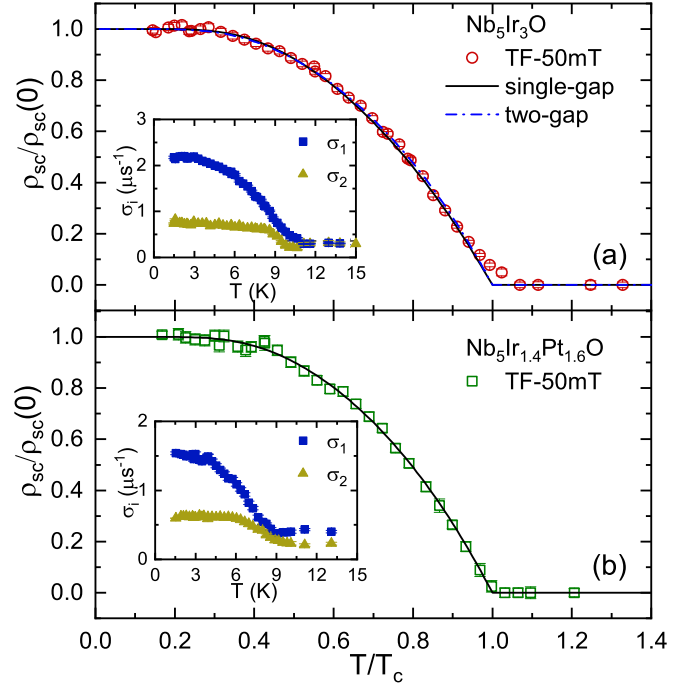}
	\caption{\label{fig:lambda}Superfluid density vs.\ reduced temperature $T/T_c$, as 
	determined from TF-$\mu$SR measurements in an applied magnetic field 
	of 50\,mT for Nb$_5$Ir$_3$O (a) and Nb$_5$Ir$_{1.4}$Pt$_{1.6}$O (b). The insets show the temperature dependence 
	of the muon-spin relaxation rate $\sigma_i(T)$. 
	The lines through the data represent fits to a fully-gapped $s$-wave 
	model with either a single gap (solid) or two gaps (dash-dotted). The goodness of fit are $\chi^2_\mathrm{r}$ = 3.8 (single-gap model) and 3.2 (two-gap model), respectively.}
\end{figure}

To investigate the superconducting properties of Nb$_5$\-Ir$_{3-x}$\-Pt$_{x}$O 
at a microscopic level, TF-$\mu$SR measurements were 
systematically carried out, covering both the normal and superconducting states. 
To track the additional field-distribution broadening due to the FLL in the mixed state, a magnetic field of 50\,mT was applied in the normal state, before cooling the sample below $T_c$. 
The TF-$\mu$SR time spetra were collected at various temperatures upon warming after the FC-protocol. 
Figure~\ref{fig:TF-muSR_T}(a) shows two representative TF-$\mu$SR spectra of Nb$_5$Ir$_3$O collected at 1.6\,K (i.e., below $T_c$) and 14\,K (above $T_c$). In the normal state, the spectra show a relatively weak damping, reflecting a uniform field distribution.  
The enhanced depolarization rate in the superconducting state is attributed to the inhomogeneous field distribution due to the FLL, causing an additional 
field broadening 
in the mixed state. Such broadening is clearly demonstrated in Fig.~\ref{fig:TF-muSR_T}(b), where the fast-Fourier-transform 
(FFT) spectra of the corresponding TF-$\mu$SR data in Fig.~\ref{fig:TF-muSR_T}(a)
are shown. To account for the asymmetric field distribution in the superconducting state, the $\mu$SR spectra were 
modeled by the following expression: 
\begin{equation}
\label{eq:TF_muSR}
A_\mathrm{TF}(t) = \sum\limits_{i=1}^n A_i \cos(\gamma_{\mu} B_i t + \phi) e^{- \sigma_i^2 t^2/2} +
A_\mathrm{bg} \cos(\gamma_{\mu} B_\mathrm{bg} t + \phi).
\end{equation}
Here $A_i$ (98\%) and $A_\mathrm{bg}$(2\%) represent the initial muon-spin asymmetries for muons implanted in the sample and sample holder, respectively, 
with the latter not undergoing any depolarization. $B_i$ and $B_\mathrm{bg}$ are the local fields sensed by implanted muons in the sample and sample holder, $\gamma_{\mu} = 2\pi \times 135.53$\,MHz/T is the muon gyromagnetic ratio, $\phi$ is the shared initial phase, and $\sigma_i$ is the Gaussian relaxation rate of the $i$th component.
More than one oscillation is required to describe the TF-$\mu$SR spectra of Nb$_5$Ir$_{3-x}$Pt$_{x}$O samples.
As illustrated in Fig.~\ref{fig:TF-muSR_T}(b), at 1.6\,K,
two broad peaks can be clearly seen, below and above the applied magnetic field (50\,mT). 
The solid lines in Figs.~\ref{fig:TF-muSR_T}(a)-(b) represent fits to Eq.~\eqref{eq:TF_muSR} with $n = 2$. 
Below $T_c$, a diamagnetic field shift appears in both samples 
[see Fig.~\ref{fig:TF-muSR_T}(c)]. 
The relaxation rate is temperature-independent and small above $T_c$, but below  $T_c$ it starts to increase due to the onset of the FLL and the increased superfluid density (see insets in Fig.~\ref{fig:lambda}).

%
%

In the case of multi-component oscillations, 
the first-term in Eq.~\eqref{eq:TF_muSR} describes the field distribution 
as a sum of $n$ Gaussian relaxations (here $n = 2$)~\cite{Maisuradze2009}:
\begin{equation}
\label{eq:TF_muSR_2}
P(B) = \gamma_{\mu} \sum\limits_{i=1}^2 \frac{A_i}{\sigma_i} \mathrm{exp}\left[-\frac{\gamma_{\mu}^2(B-B_i)^2}{2\sigma_i^2}\right].
\end{equation}
The first- and second moments of the field distribution can be calculated by: 
\begin{equation}
\begin{aligned}
\label{eq:1st_moment}
&	\langle B \rangle =  \sum\limits_{i=1}^2 \frac{A_i B_i}{A_\mathrm{tot}},\quad \mathrm{and} \\
%
%
& \langle B^2 \rangle = \frac{\sigma_\mathrm{eff}^2}{\gamma_\mu^2} = \sum\limits_{i=1}^2 \frac{A_i}{A_\mathrm{tot}}\left[\frac{\sigma_i^2}{\gamma_{\mu}^2} + \left(B_i - \langle B \rangle\right)^2\right],
\end{aligned}
\end{equation}
where $A_\mathrm{tot} = A_1 + A_2$. 
The superconducting Gaussian 
relaxation rate $\sigma_\mathrm{sc}$ can be 
extracted by subtracting the nuclear contribution according to 
$\sigma_\mathrm{sc} = \sqrt{\sigma_\mathrm{eff}^{2} - \sigma^{2}_\mathrm{n}}$, 
where $\sigma_\mathrm{n}$ is the nuclear relaxation rate, 
assumed 
to be temperature independent in such a narrow temperature range (see also ZF-$\mu$SR below).

Since the upper critical fields of Nb$_5$Ir$_{3-x}$Pt$_{x}$O (see Fig.~\ref{fig:Hc2_determ}) are significantly higher than the transverse field used 
in the $\mu$SR measurements (here, 50\,mT), 
the magnetic penetration depth $\lambda(T)$ and the superfluid density $\rho_{sc}(T)$ [$\propto \lambda^{-2}(T)$] can be obtained from $\sigma_\mathrm{sc}$$(T)$ according to~\cite{Barford1988,Brandt2003}:

\begin{equation}
\label{eq:sig_to_lam}
\frac{\sigma_\mathrm{sc}^2(T)}{\gamma^2_{\mu}} = 0.00371\, \frac{\Phi_0^2}{\lambda^4(T)}.
\end{equation}
The derived superfluid density normalized to the 
zero-temperature values is shown in the main panels of Fig.~\ref{fig:lambda}. 
The superfluid density is almost constant at temperatures below $T_c/3$, indicating fully-gapped SC in Nb$_5$Ir$_{3-x}$Pt$_{x}$O, in good agreement with the specific heat results (see Fig.~\ref{fig:Cp2}).

For a more quantitative insight into the SC of Nb$_5$\-Ir$_{3-x}$\-Pt$_{x}$O, 
the derived superfluid density $\rho_\mathrm{sc}(T)$ was further analyzed by using a fully-gapped $s$-wave model:
\begin{equation}
\label{eq:rhos}
\frac{\lambda^{-2}(T)}{\lambda_0^{-2}} = \frac{\rho_\mathrm{sc}(T)}{\rho_\mathrm{sc}(0)} =  1 + 2\int^{\infty}_{\Delta(T)} \frac{E}{\sqrt{E^2-\Delta^2(T)}} \frac{\partial f}{\partial E} dE.
\end{equation}
Here $f$ and $\Delta(T)$ are the Fermi and superconducting gap functions (see details in Sec.~\ref{ssec:Cp_zero}). 
The solid black lines in Fig.~\ref{fig:lambda} are fits to the above model with a single gap, which yield zero-temperature  
gap values $\Delta_0$ = 1.79(3) and 1.67(2)\,meV, and magnetic penetration depths $\lambda_0$ = 230(2) and 314(2)\,nm, for Nb$_5$Ir$_3$O and Nb$_5$Ir$_{1.4}$Pt$_{1.6}$O, respectively. 
The estimated BCS coherence length $\xi_0$ is larger than the electronic mean 
free path $l_\mathrm{e}$ (see Table~\ref{tab:parameter}), implying the samples are  
in the dirty limit. Therefore, 
the temperature-dependent superfluid density was also 
analyzed using a dirty-limit model.
\tcr{In this case, in the BCS approximation, the temperature 
dependence of the superfluid density is given by $\rho_\mathrm{sc}(T) = \frac{\Delta(T)}{\Delta_0} \mathrm{tanh} \left[\frac{\Delta(T)}{2k_\mathrm{B}T}\right]$~\cite{Tinkham1996}, }
%
%
which yields a gap
value of 1.61(3) and 1.51(3)\,meV, slightly smaller than the clean-limit
value, but still larger than the weak-coupling BCS value. 
In the Nb$_5$Ir$_3$O case, to compare the zero-field electronic specific heat results~\cite{Zhang2017,Wang2019} with those from $\mu$SR, the superfluid density  was also analyzed using a two-gap model~\cite{Carrington2003,Nieder2002,Shang2019}. 
As shown by the blue dash-dotted line in Fig.~\ref{fig:lambda} (a), the two-gap model shows a slightly better agreement with the $\rho_\mathrm{sc}(T)$ data, as confirmed by the smaller $\chi^2_\mathrm{r}$ value. The derived gap values are $\Delta_0^\mathrm{s}$ = 1.34(3)\,meV (small) and $\Delta_0^\mathrm{l}$ = 1.97(3)\,meV (large), with a weight $w$ = 0.2 of the small gap. Such a small weight makes the multigap features barely visible in the $\rho_{sc}(T)$ or $C_\mathrm{e}/T$ data. 
Although the $\rho_{sc}(T)$ data can be well described by an $s$-wave model with either single- or two gaps, the latter is more consistent with 
the field-dependent electronic specific heat coefficient (Fig.~\ref{fig:Cp3}), the superconducting Gaussian relaxation rate (Fig.~\ref{fig:lambda2}), 
the upper critical field (Fig.~\ref{fig:Hc2_twoband}), and the electronic band structure calculations~\cite{Zhang2017}.

\subsection{\label{ssec:twoband} Evidence of multigap superconductivity in Nb$_5$Ir$_3$O}

\begin{figure}[th]
	\centering
	\includegraphics[width=0.48\textwidth,angle=0]{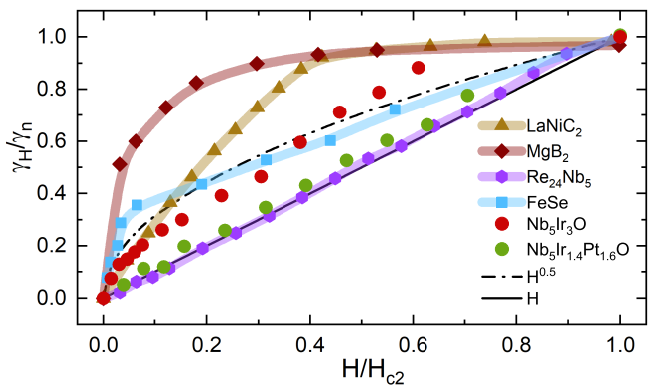}
	\vspace{-2ex}%
	\caption{\label{fig:Cp3}The normalized 
		specific heat coefficient $\gamma_\mathrm{H}$/$\gamma_\mathrm{n}$ 
		vs.\ the reduced magnetic field $H/H_\mathrm{c2}(0)$ for Nb$_5$Ir$_{3-x}$Pt$_x$O. 
		$\gamma_\mathrm{H}$ is obtained as the linear extrapolation of $C/T$ vs.\ $T^2$ (in the superconducting phase) to zero temperature. 
	    The solid line indicates a linear dependence, as predicted for a single-gap $s$-wave 
		gap structure, the dash-dotted line represents the dependence expected for an anisotropic gap or a gap with nodes, e.g., $d$-wave. The data of the reference samples are adopted from Refs.~\onlinecite{Chen2013,Chen2017,Bouquet2001a,TianReNb2018}.} 
\end{figure}
%

\begin{figure}[!thp]
	\centering
	\includegraphics[width=0.49\textwidth,angle= 0]{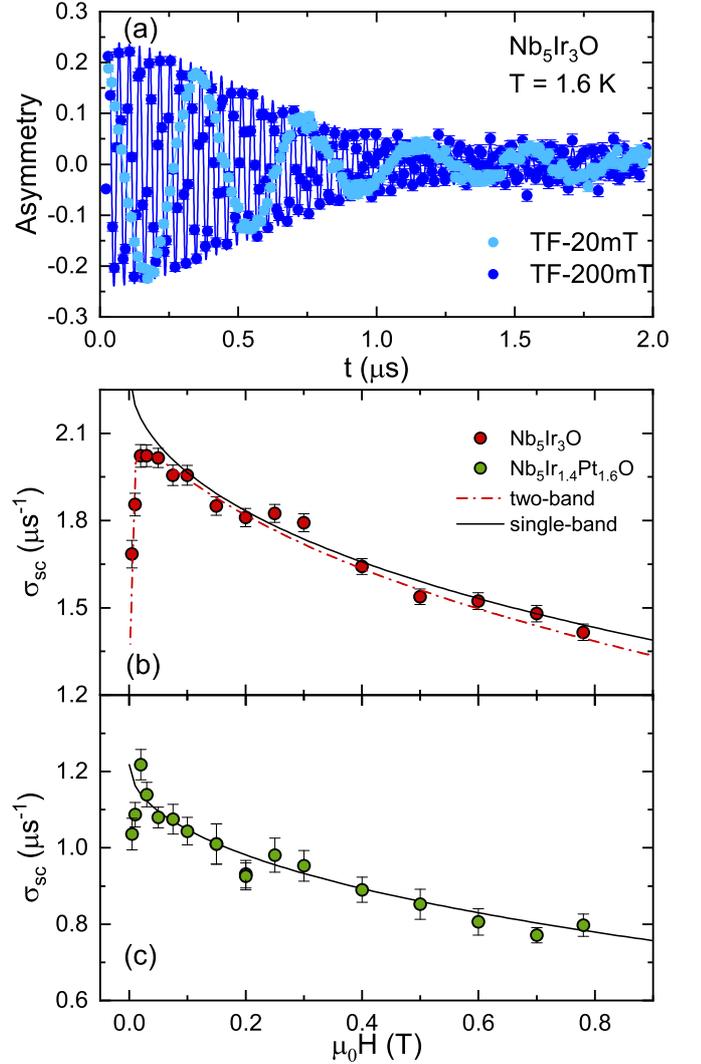}
	\caption{\label{fig:lambda2}(a) TF-$\mu$SR time spectra for Nb$_5$Ir$_3$O measured at $T = 1.5$\,K (superconducting state) in a field of 20 and 200\,mT. \tcr{Nb$_5$Ir$_{1.4}$Pt$_{1.6}$O, too, shows a similar behavior. Field-dependent superconducting Gaussian relaxation rate 
		$\sigma_\mathrm{sc}(H)$ for Nb$_5$Ir$_3$O (b) and Nb$_5$Ir$_{1.4}$Pt$_{1.6}$O (c). The dash-dotted and solid lines 
		represent fits to two-band and single-band models, respectively. The goodness of fit are $\chi^2_\mathrm{r}$ = 1.17 (two-band model) and 2.65 (single-band model) for Nb$_5$Ir$_3$O, and $\chi^2_\mathrm{r}$ = 1.15
		for Nb$_5$Ir$_{1.4}$Pt$_{1.6}$O (single-band model). The data for $H < H_\mathrm{c1}$ are excluded to evaluate the $\chi^2_\mathrm{r}$ value.}}
\end{figure}

The multigap nature  
of Nb$_5$Ir$_3$O superconductivity can be further inferred  
from the field-dependent electronic specific heat coefficient $\gamma_\mathrm{H}(H)$. 
The normalized values $\gamma_\mathrm{H}/\gamma_\mathrm{n}$ vs.\ 
the reduced magnetic field $H/H_\mathrm{c2}(0)$ are shown 
in Fig.~\ref{fig:Cp3} (here $\gamma_\mathrm{n}$ is the zero-field normal-state value).  
For Nb$_5$Ir$_3$O, due to its multigap nature, it is difficult to describe the field dependence with a simple formula. 
$\gamma_\mathrm{H}(H)$ clearly deviates 
from the linear field dependence (solid line) expected for 
single gap fully-gapped superconductors~\cite{Caroli1964}, 
or from the square-root dependence $\sqrt{H}$ (dash-dotted line), 
expected for nodal superconductors~\cite{Volovik1993,Wen2004}. 
Different from the case of Nb$_5$Ir$_3$O, but 
similar to Re$_{24}$Nb$_{5}$~\cite{TianReNb2018}, $\gamma_\mathrm{H}(H)$ of Nb$_5$Ir$_{1.4}$Pt$_{1.6}$O is practically linear in field, more consistent with a single gap SC.
Nb$_5$Ir$_3$O instead exhibits features similar to 
other well studied multigap superconductors, as e.g., LaNiC$_2$~\cite{Chen2013}, 
FeSe~\cite{Chen2017}, and MgB$_2$~\cite{Bouquet2001a}, although the slopes 
of $\gamma_\mathrm{H}(H)$ close to zero field are different, 
reflecting the different magnitudes and weights of the smaller gap.

\begin{figure}[htp]
	\centering
	\includegraphics[width=0.47\textwidth,angle= 0]{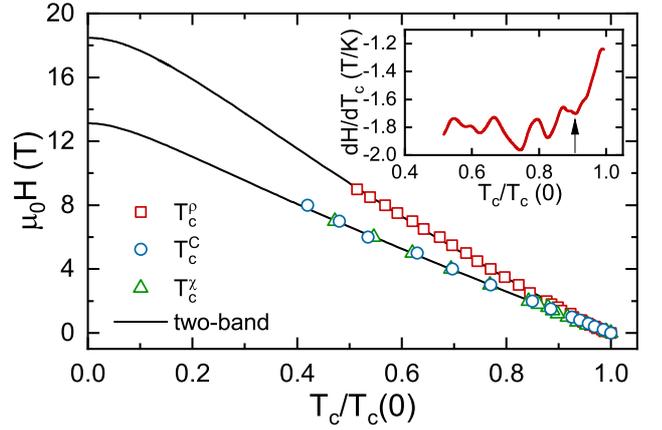}
	\caption{\label{fig:Hc2_twoband}Upper critical field $\mu_{0}H_{c2}$ vs.\ 
		reduced transition temperature $T_c/T_c(0)$ for Nb$_5$Ir$_3$O [same as in Fig.~\ref{fig:Hc2_determ}(a)]. The solid lines are fits using a two-band model. \tcr{Inset: Temperature derivative of the upper critical field,  as determined from the electrical resistivity.}}
\end{figure}

To get further insight into the multigap SC of Nb$_5$Ir$_3$O, TF-$\mu$SR measurements were performed in different magnetic fields up to 780\,mT
at base temperature (1.6\,K) in both samples. 
Figure~\ref{fig:lambda2}(a) shows the TF-$\mu$SR 
spectra of Nb$_5$Ir$_3$O, collected at 20 and 200\,mT, 
with the  spectra in other applied fields and in Nb$_5$Ir$_{1.4}$Pt$_{1.6}$ showing similar features. 
The spectra were analyzed using the same model as described in 
Eq.~\eqref{eq:TF_muSR}, and the resulting superconducting Gaussian 
relaxation rates $\sigma_\mathrm{sc}$ versus the applied magnetic field 
are summarized in 
Fig.~\ref{fig:lambda2}(b) (Nb$_5$Ir$_3$O) and Fig.~\ref{fig:lambda2}(c) (Nb$_5$Ir$_{1.4}$Pt$_{1.6}$). 
$\sigma_\mathrm{sc}(H)$ was analyzed using both a single- and a two-band model. In the latter case, each band is 
characterized by its own superconducting coherence length 
[i.e., $\xi_1(0)$ and $\xi_2(0)$] and a weight $w$ accounting for the contribution of the second band [$\xi_2(0)$] to the total superfluid density, similar to the two-gap model in Fig.~\ref{fig:lambda}(a).
The details of the single- and two-band models can be found in Refs.~\onlinecite{Serventi2004,Khasanov2014}. By fixing $w = 0.2$ and $\xi_1(0) = 5.4$\,nm, 
as estimated from the analysis of $\rho_{sc}(T)$ (Fig.~\ref{fig:lambda}) 
and of the upper critical field (Fig.~\ref{fig:Hc2_determ}), 
we obtain the dash-dotted line fit in Fig.~\ref{fig:lambda2}(b), 
which provides 
$\lambda_0$ = 222(3)\,nm and $\xi_2(0)$ = 14(1)\,nm. The derived $\lambda_0$ is consistent with the value (230\,nm) estimated from the analysis of $\rho_{sc}(T)$ in Fig.~\ref{fig:lambda}(a).  
The upper critical field of 1.7(2)\,T calculated from the coherence length of the second band $\xi_2(0)$ is also in good agreement with the field values where $\gamma_\mathrm{H}(H)$ and $\mu_0H_\mathrm{c2}(T)$ change their slope, as shown by the arrows in Fig.~\ref{fig:rho}(a) and the inset in Fig.~\ref{fig:Hc2_twoband}. 
\tcr{While the single-band model reproduces very well the $\sigma_\mathrm{sc}(H)$ data for Nb$_5$Ir$_{1.4}$Pt$_{1.6}$ [see Fig~\ref{fig:lambda2}(c)], it is less satisfactory for Nb$_5$Ir$_3$O, where a two-band model is required to obtain 
a similarly small $\chi^2_\mathrm{r}$ value [see Fig~\ref{fig:lambda2}(b)].}

A positive curvature of the upper critical field near $T_c$ is considered a 
typical feature of multiband superconductors, as e.g., MgB$_2$~\cite{Muller2001,Gurevich2004}.  
It reflects the gradual suppression of the small superconducting gap
with increasing magnetic field. 
Indeed, in Nb$_5$Ir$_3$O, $\mu_0 H_{c2}(T)$ exhibits a clear kink 
close to 1.5\,T [see arrow in Fig.~\ref{fig:rho}(a)], which coincides 
with the field value which suppresses the small superconducting gap. 
This is reflected also in the derivative of $\mu_0 H_{c2}(T)$ with 
respect to temperature (see arrow in the inset of Fig.~\ref{fig:Hc2_twoband}), 
Also $\gamma_\mathrm{H}(H)$ changes its slope near this critical field (close to 0.1 in Fig.~\ref{fig:Cp3}).
Since, in Nb$_5$Ir$_{1.4}$Pt$_{1.6}$O, the small gap is already 
suppressed by Pt substitution, it exhibits a linear field dependence 
of $\gamma_\mathrm{H}(H)$ (Fig.~\ref{fig:Cp3}), 
consistent with a fully-gapped SC with a single gap.    
As shown in Fig.~\ref{fig:Hc2_twoband}, 
the upper critical field of Nb$_5$Ir$_3$O was also analyzed by a two-band model~\cite{Gurevich2011}, from which 
we estimate the upper critical field values $\mu_0 H_{c2}(0)$ = 13.1(1) [from $C(T,H)$ or $M(T,H)$] and 18.5(1)\,T [from $\rho(T,H)$], both consistent with the GL model in Fig.~\ref{fig:Hc2_determ}(a). 

%
%

\subsection{\label{ssec:Dis} Uemura plot}
%

\begin{figure}[ht]
	\centering
	\includegraphics[width = 0.46\textwidth]{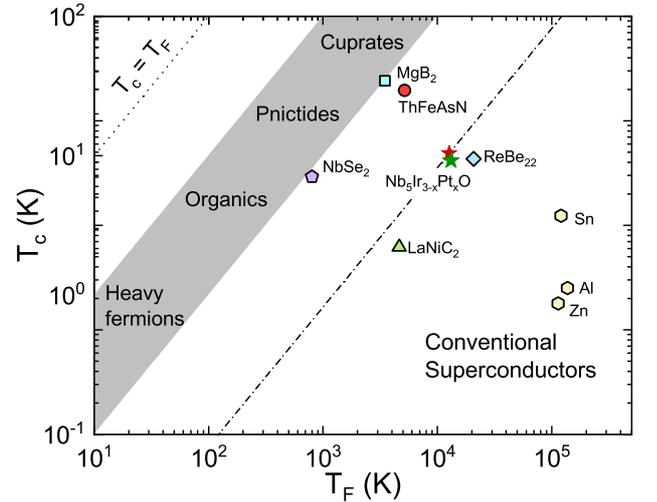}
	\caption{\label{fig:uemura}Uemura plot showing
		$T_c$ against the effective Fermi temperature $T_\mathrm{F}$ for various superconductors.   
		The shaded region, with  $1/100 < T_c/T_\mathrm{F}<1/10$,  indicates the band of unconventional 
		superconductors, such as heavy fermions,  organic superconductors, iron pnictides and cuprates. 
		The dotted line corresponds to $T_c = T_\mathrm{F}$, while the dash-dotted 
		line indicates $T_c/T_\mathrm{F} = 8.2 \times 10^{-4}$ for Nb$_5$Ir$_3$O. 
		The data of the reference samples were adopted from Refs.~\onlinecite{Shang2019,Shiroka2017,Uemura1991,TianReNb2018,Uemura2006}.}
\end{figure}

According to the ratio of $T_c$ to the effective Fermi temperature $T_\mathrm{F}$, 
the different classes of superconductors can be classified  following the so-called
Uemura plot~\cite{Uemura1991}. As seen in 
Fig.~\ref{fig:uemura}, conventional BCS superconductors exhibit 
$T_c/T_\mathrm{F} <10^{-3}$, here exemplified by the elemental Sn, Al, and Zn superconductors. 
In contrast, as indicated by the shadowed region,  
several types of unconventional superconductors, including 
heavy-fermions, organic superconductors, iron pnictides and cuprates, 
all lie within a $10^{-2} < T_c/T_\mathrm{F} <10^{-1}$ band. 
Between these two categories are located several 
multigap superconductors, as e.g., LaNiC$_2$, NbSe$_2$
and MgB$_2$. According to the 
superconducting parameters obtained from the measurements presented here
(see details in Table.~\ref{tab:parameter}), the calculated $T_c/T_\mathrm{F}$ 
values for Nb$_5$Ir$_3$O and Nb$_5$Ir$_{1.4}$Pt$_{1.6}$O are $\sim$ 8.2-8.4 $\times 10^{-4}$ 
(see star symbols in Fig.~\ref{fig:uemura}).
Although there is no evidence for them to be classified as unconventional 
superconductors, the Nb$_5$Ir$_{3-x}$Pt$_x$O family is clearly far off 
the conventional superconductors 
and it shows similar ratios to other multigap superconductors, such as  
LaNiC$_2$ and ReBe$_{22}$ (both located near the dash-dotted line). 

%
\begin{table}[!bht]
	\centering
	\caption{Normal- and superconducting state properties of Nb$_5$Ir$_3$O and Nb$_5$Ir$_{1.4}$Pt$_{1.6}$O, as 
		determined from electrical resistivity, magnetic susceptibility, 
		specific heat, and $\mu$SR measurements. The London penetration 
		depth $\lambda_\mathrm{L}$, the effective mass $m^{\star}$, 
		carrier density $n_s$, BCS coherence length $\xi_0$, electronic 
		mean-free path $l_e$, Fermi velocity $v_\mathrm{F}$, and effective Fermi 
		temperature $T_\mathrm{F}$ were estimated following equations 
		in Ref.~\onlinecite{Barker2018}.
		\label{tab:parameter}}
	\begin{ruledtabular}
		\begin{tabular}{lccc}
			Property                               & Unit            & Nb$_5$Ir$_3$O      & Nb$_5$Ir$_{1.4}$Pt$_{1.6}$O     \\ \hline
			$T_c$\footnotemark[1]                  & K               & 10.5(1)            & 9.1(1)        \rule{0pt}{2.6ex} \\		
			$\mu_0H_{c1}$                          & mT              & 11.5(2)            & 8.0(1)      \\
			$\mu_0H_{c1}$$^{\mu\mathrm{SR}}$       & mT              & 13.3(1)            & 7.7(1)      \\
			$\mu_0H_{c2}$$^\rho$                   & T               & 15.5(1)            & 12.7(1)       \\
			$\mu_0H_{c2}$$^{C}$                    & T               & 11.2(1)            & 12.7(1)     \\[2mm]		
			$\gamma_n$                             & mJ/mol-K$^2$    & 37(5)              & 42(6)       \\
			$\Theta_\mathrm{D}^\mathrm{C}$         & K               & 379(30)            &368 (30)     \\
			$\lambda_\mathrm{ep}$                  & ---             & 0.8(2)           &0.73(25)     \\
			$N(\epsilon_\mathrm{F})$               & states/eV-f.u.  & 16(2)               &18(2)         \\[2mm]
			$\Delta_0$$^{\mu\mathrm{SR}}$(clean)   & meV             & 1.79(3)\footnotemark[2]             & 1.67(2) \\
			$\Delta_0$$^{\mu\mathrm{SR}}$(dirty)   & meV             & 1.61(3)            & 1.51(3)  \\  
			$\Delta_0^{C}$                         & meV             & 1.89(2)            & 1.53(1)  \\
			$\Delta C/\gamma_\mathrm{n}T_c$        & ---             & 2.24(7)            & 1.50(5)  \\[2mm]
			$\lambda_0^{\mu\mathrm{SR}}$           & nm              & 230(2)             & 314(2)  \\
			$\lambda_\mathrm{GL}$                  & nm              & 249(3)             & 308(3)  \\
			$\lambda_\mathrm{L}$                   & nm              & 73(5)              & 81(9)   \\
			$\xi(0)$                               & nm              & 5.4(1)             & 5.1(1)  \\
			$\kappa$                               & ---             & 50(3)              & 60(3)   \\[2mm]		
			$m^{\star}$                            & $m_e$           & 8.0(8)             & 9.2(4)  \\
			$n_\mathrm{s}$                         & 10$^{28}$\,m$^{-3}$ & 4.2(5)         & 3.9(7)  \\
			$\xi_0$                                & nm              & 18(1)              & 16(2)   \\
			$l_e$                                  & nm              & 2.1(1)             & 1.2(1)  \\
			$\xi_0$/$l_e$                          & ---             & 9(1)               & 14(2)   \\ 
			$v_\mathrm{F}$                         & 10$^5$\,ms$^{-1}$ & 1.5(1)           & 1.3(1)  \\
			$T_\mathrm{F}$                         & 10$^4$\,K         & 1.27(5)          & 1.1(1)  \\
		\end{tabular}
		\footnotetext[1]{Similar values were determined via electrical resistivity, magnetic sus\-cep\-ti\-bi\-li\-ty, and specific heat measurements.}.
		\footnotetext[2]{The two-gap model provides an averaged gap value of 1.84(3)\,meV.}	
	\end{ruledtabular}
\end{table}

\subsection{\label{ssec:ZF_muSR}Zero-field \texorpdfstring{$\mu$SR}{MuSR}}
%
\begin{figure}[ht]
	\centering
	\includegraphics[width=0.49\textwidth,angle=0]{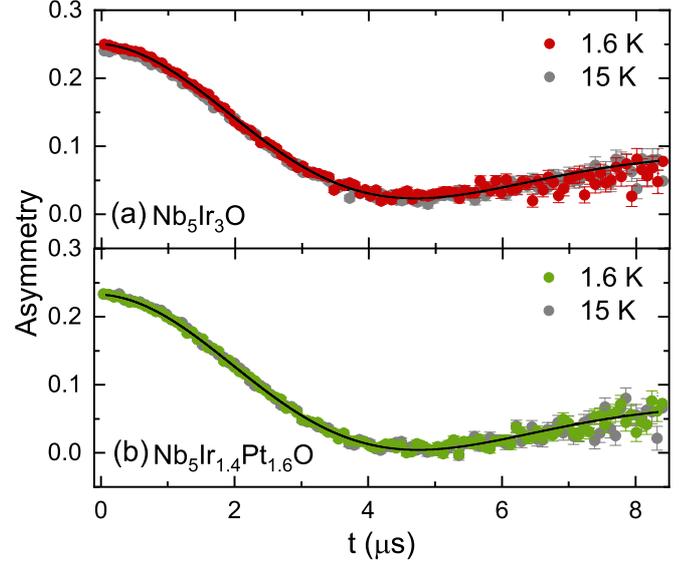}
	\vspace{-2ex}%
	\caption{\label{fig:ZF_muSR}ZF-$\mu$SR spectra for \NIO~(a) and \NIPO~(b)
	   in the superconducting (1.6\,K) and the normal state (15\,K).   
		Solid lines are fits to the equation described in the text. 
		None of the datasets shows clear changes with temperature.}
\end{figure}

%
ZF-$\mu$SR measurements were also performed 
in the normal- and superconducting states to search for possible magnetism or 
time-reversal symmetry breaking in the superconducting state of Nb$_5$Ir$_{3-x}$Pt$_x$O.
Representative ZF-$\mu$SR spectra collected above (15\,K)
and below $T_c$ (1.6\,K) for Nb$_5$Ir$_{3-x}$Pt$_x$O are shown in Fig.~\ref{fig:ZF_muSR}. 
Neither coherent oscillations nor fast damping could be identified 
at either temperature, implying the lack of any magnetic order/fluctuations in  
Nb$_5$Ir$_{3-x}$Pt$_x$O. Therefore, in absence of an applied magnetic field, the weak muon-spin relaxation is mainly determined by the randomly oriented nuclear moments, which can be modeled by a Gaussian Kubo-Toyabe relaxation function $G_\mathrm{KT} = [\frac{1}{3} + \frac{2}{3}(1 -\sigma_\mathrm{ZF}^{2}t^{2})\,\mathrm{e}^{-\frac{\sigma_\mathrm{ZF}^{2}t^{2}}{2}}] $~\cite{Kubo1967,Yaouanc2011}. 
The solid lines in Fig~\ref{fig:ZF_muSR} represent fits to the data by considering an additional zero-field Lorentzian relaxation $\Lambda$, i.e., $A_\mathrm{ZF} = A_\mathrm{s} G_\mathrm{KT} \mathrm{e}^{-\Lambda t} + A_\mathrm{bg}$. 
Here $A_\mathrm{s}$ and $A_\mathrm{bg}$ are the same as in the TF-$\mu$SR case [see Eq.~(\ref{eq:TF_muSR})]. 
The strong Gaussian relaxation rates reflect the large nuclear moments in Nb$_5$Ir$_{3-x}$Pt$_x$O, mostly determined by the Nb nuclear moments.
In both the normal- and the superconducting states, the relaxations 
are almost identical, as demonstrated by the practically overlapping ZF-$\mu$SR spectra above and below $T_c$. 
This lack of evidence for an additional $\mu$SR relaxation below $T_c$ excludes a possible time-reversal symmetry breaking 
in the superconducting state of Nb$_5$Ir$_{3-x}$Pt$_x$O.

\vspace{7pt}
\section{\label{ssec:Sum} Conclusion}
To summarize, we investigated the superconducting properties of 
Nb$_{5}$Ir$_{3-x}$Pt$_x$O (for $x = 0$ and 1.6)
by means of electrical resistivity, magnetization, specific heat, and $\mu$SR measurements.  
Nb$_5$Ir$_3$O and Nb$_5$Ir$_{1.4}$Pt$_{1.6}$O exhibit bulk $T_c$ at 10.5 and 9.1\,K, respectively. 
The temperature dependence of the zero-field electronic specific heat 
and superfluid density reveal a nodeless SC in Nb$_{5}$Ir$_{3-x}$Pt$_x$O, well 
described by an isotropic $s$-wave model.
Nb$_5$Ir$_3$O, instead, turns out to be a multigap superconductor, 
as demonstrated by the field dependence of the 
electronic specific heat coefficient, the superconducting Gaussian relaxation, 
and the temperature dependence of its upper critical field. 
Upon Pt substitution, the small superconducting gap is suppressed 
and Nb$_5$Ir$_{1.4}$Pt$_{1.6}$O shows typical features of single gap SC, 
hence indicating a crossover from multiple- to single gap 
SC in the Nb$_{5}$Ir$_{3-x}$Pt$_x$O family.
Finally, the lack of spontaneous magnetic fields below $T_c$ 
indicates that the time-reversal symmetry is preserved in Nb$_{5}$Ir$_{3-x}$Pt$_x$O superconductors. 

\begin{acknowledgments} 
The authors thank R.\ Khassanov for fruitful discussions and acknowledge the assistance from the S$\mu$S beamline scientists. 
This work was supported by the Schwei\-ze\-rische Na\-ti\-o\-nal\-fonds 
zur F\"{o}r\-de\-rung der Wis\-sen\-schaft\-lich\-en For\-schung, 
SNF (Grants No.\ 200021\_169455 and 206021\_139082). 
\end{acknowledgments}
%

\bibliography{NbIrO_bib}

\end{document}